\documentclass[10pt, a4paper]{article}
\pdfoutput=1 
\usepackage{jcappub}
\usepackage{graphicx}
\usepackage{amsmath,amssymb}
\usepackage{bm}
\usepackage{color}
\usepackage{tensor}
\usepackage{hyperref}
\usepackage{slashed}
\usepackage{multirow}
\usepackage{natbib}
\usepackage{braket} 

\newcommand{\be}{\begin{equation}}
\newcommand{\ee}{\end{equation}}
\def\ba{\begin{eqnarray}}
\def\ea{\end{eqnarray}}
\def \bse {\begin{subequations} \begin{eqnarray}}
\def \ese {\end{eqnarray} \end{subequations}}
\def\bald{\begin{aligned}}
\def\eald{\end{aligned}}

\def \dd  {{\rm d}}
\def \l {\left}
\def \r {\right}
\newcommand{\e}[1]{{\rm e}^{#1}}

\newcommand{\moy}[1]{\langle #1 \rangle}

\newcommand{\pder}[2]{\dfrac{\partial #1}{\partial #2}}
\newcommand{\per}{\, .}
\newcommand{\com}{\, ,}
\newcommand{\eref}[1]{Eq.~(\ref{#1})}

\def\L{{\scriptscriptstyle L}}
\def\R{{\scriptscriptstyle R}}

\def\D{{\scriptscriptstyle D}}
\def\yql{y_{\scriptscriptstyle{q}}}
\def\yll{y_{\scriptscriptstyle{\ell}}}
\def\yur{y_{u_{\scriptscriptstyle R}}}
\def\ydr{y_{d_{\scriptscriptstyle R}}}
\def\yer{y_{e_{\scriptscriptstyle R}}}
\def\yli{y_{L_{\scriptscriptstyle i}}}
\def\yri{y_{R_{\scriptscriptstyle i}}}
\def\alphay{\alpha_{\scriptscriptstyle \rm y}}
\def\alphaw{\alpha_{\scriptscriptstyle \rm w}}
\def\alphas{\alpha_{\scriptscriptstyle \rm s}}
\def\Nc{N_{\rm c}}
\def\Nw{N_{\rm w}}
\def\NgSM{N_{\rm g}^{\rm SM}}
\def\NgDM{N_{\rm g}^{\rm DM}}
\def\muql{\mu_{{\scriptscriptstyle q}}}
\def\mull{\mu_{{\scriptscriptstyle \ell}}}
\def\muur{\mu_{{\scriptscriptstyle u}_{\scriptscriptstyle R}}}
\def\mudr{\mu_{{\scriptscriptstyle d}_{\scriptscriptstyle R}}}
\def\muer{\mu_{{\scriptscriptstyle e}_{\scriptscriptstyle R}}}
\def\muL{\mu_{L}}
\def\muR{\mu_{R}}
\def\edotb{{\bm E} \cdot {\bm B}}
\def\gammay{\gamma_{\rm \scriptscriptstyle y}}
\def\gammaws{\gamma_{\rm \scriptscriptstyle w}}
\def\gammass{\gamma_{\rm \scriptscriptstyle s}}
\def\gammaw{\Gamma_{\rm w}}
\def\gammas{\Gamma_{\rm s}}
\def\etaql{\eta_{{\scriptstyle q}}}
\def\etall{\eta_{{\scriptstyle \ell}}}
\def\etaur{\eta_{u_{\scriptscriptstyle R}}}
\def\etadr{\eta_{d_{\scriptscriptstyle R}}}
\def\etaer{\eta_{e_{\scriptscriptstyle R}}}
\def\etaL{\eta_{L}}
\def\etaR{\eta_{R}}
\def\cy{C_{\rm y}}
\def\cw{C_{\rm w}}
\def\cs{C_{\rm s}}
\def\hinf{H_{\rm inf}}

\def\Mpl{M_{\rm pl}}
\def\Hinf{H_{\text{inf}}}

\newcommand{\uy}{\rm{U(1)}_{\scriptscriptstyle Y}}
\newcommand{\suL}{\rm{SU(2)}_{\scriptscriptstyle L}}
\newcommand{\suc}{\rm{SU(3)}_{\scriptscriptstyle c}}
\newcommand{\ud}{\rm{U(1)}_{\scriptscriptstyle D}}
\newcommand{\dLo}{D_{\scriptscriptstyle L_1}}
\newcommand{\dLt}{D_{\scriptscriptstyle L_2}}
\newcommand{\qd}{Q_{\scriptscriptstyle D}}
\newcommand{\gd}{g_{\scriptscriptstyle D}}
\newcommand{\yLo}{y_{\scriptscriptstyle L_1}}
\newcommand{\yLt}{y_{\scriptscriptstyle L_2}}
\newcommand{\yXLo}{y_{\scriptscriptstyle X L_1}}
\newcommand{\yXLt}{y_{\scriptscriptstyle X L_2}}
\newcommand{\yXLi}{y_{\scriptscriptstyle X L_i}}

\title{\LARGE Asymmetric Dark Matter and Baryogenesis from Pseudoscalar Inflation}

\date{\today}

\author[1]{Yann Cado,}
\author[1,2]{Eray Sabancilar} 

\affiliation[1]{Laboratory of Particle Physics and Cosmology, Institute of Physics,
Ecole Polytechnique F\'ed\'erale de Lausanne, CH-1015 Lausanne, Switzerland}
\affiliation[2]{Department of Physics, Bielefeld University, D-33615, Bielefeld, Germany}
\emailAdd{yann.cado@epfl.ch}
\emailAdd{eray.sabancilar@epfl.ch}

\abstract{
We show that both the baryon asymmetry of the Universe and the dark matter abundance can be explained within a single framework that makes use of maximally helical hypermagnetic fields produced during pseudoscalar inflation and the chiral anomaly in the Standard Model. We consider a minimal asymmetric dark matter model free from anomalies and constraints. We find that the observed baryon and the dark matter abundances are achieved for a wide range of inflationary parameters, and the dark matter mass ranges between 7-15 GeV. The novelty of our mechanism stems from the fact that the same source of CP violation occurring during inflation explains both baryonic and dark matter in the Universe with two inflationary parameters, hence addressing all the initial condition problems in an economical way.
}

\keywords{asymmetric dark matter, baryogenesis, CP violation, chiral anomaly, pseudoscalar inflation, magnetic helicity}

\begin{document}
\maketitle

\section{Introduction}
\label{sec:introduction}

The curious coincidence between the observed baryon and dark matter abundances lead to the so called asymmetric dark matter (ADM) paradigm, where the dark sector mimics the baryonic one by exhibiting an asymmetry in its abundance of particles over its antiparticles (see e.g., reviews \cite{Petraki:2013wwa,Zurek:2013wia}). The basic idea behind the asymmetric dark matter scenario is that the same source of $\cal CP$ violation that leads to the baryogenesis also feeds into the dark sector, and hence similar abundances are achieved in both. Typically, in such models the dark matter candidate has a mass not so far from the tens of GeV to a few GeV unless there is a huge suppression or enhancement factor for the transfer of the asymmetry. In that sense it is a quite predictive top down approach. 

On the cosmological side, it is still not clear what the source of baryon asymmetry of the Universe (BAU) is and at which epoch it occurred. There are vast ways of generating the BAU, but there are only a few testable models (see e.g., $\nu$MSM \cite{Asaka:2005an,Asaka:2005pn}) due to having too many parameters and/or not being in reach for accelerator experiments or cosmological observations. 

It has recently been pointed out that $\cal CP$ violation that occurs during inflation via a coupling of an inflaton to the hypercharge gauge fields via a dimension 5 operator of the form $(\alpha/f)\Phi F_{\mu\nu}{\tilde F}^{\mu\nu}$ leads to a successful baryogenesis\footnote{Ref.~\cite{Alexander:2004us} also considered a somewhat similar mechanism using the coupling of inflaton to gravity and by making use of gravitational anomaly.} \cite{Anber:2015yca}. The basic idea is that during inflation there is a non-perturbative production of gauge fields with high occupation numbers leading to coherent maximally helical hypermagnetic fields \cite{Anber:2006xt}, which in turn sources the well known chiral anomaly in the Standard Model (SM) producing an asymmetry in the SM particle species. The model only depends on two parameters, namely the scale of inflation, $\Hinf$ and the coupling of the inflaton to the hypercharge gauge fields $\alpha$. With these basic ingredients, all that is needed to produce the required asymmetries is the SM physics, namely the chiral anomaly. In this work, we generalize this framework to include the generation of asymmetric dark matter and report on a relation between the $\cal CP$ violation that occurs during pseudoscalar inflation and the observed baryon and dark matter abundances in the Universe. Hence, we propose a mechanism that solves all the initial condition problems including the baryon and dark matter abundances.

This paper is organized as follows. In Sec.~\ref{sec:adm}, we introduce our minimal ADM field content and interactions. We review the chiral anomaly in the SM in Sec.~\ref{sec:chiral anomaly}. We discuss the generation and evolution of maximally helical hypermagnetic fields during pseudoscalar inflation in Sec.~\ref{sec:field generation} and Sec.~\ref{sec:field evolution}, respectively, and calculate the rate of change of hypermagnetic helicity that feeds into the chiral anomaly in Sec.~\ref{sec:helicity}. We introduce the asymmetry parameters and the associated kinetic equations governing their evolution in Sec.~\ref{sec:boltzmann}. Our results appear in Sec.~\ref{sec:results}. We discuss various mechanisms for the annihilation of the symmetric part of the ADM in Sec.~\ref{sec:annihilation}. We conclude with a summary of our results and a discussion in Sec.~\ref{sec:conclusion}. 

We use natural units $c=1$, $\hbar =1$ and set the Boltzmann constant $k_B =1$. We parametrize the flat Friedman-Robertson-Walker (FRW) metric as $\dd s^2 = a(\tau)^2 (\dd \tau^2 - \dd {\bm x}^2)$, where $a$ is the scale factor, $\tau$ is the conformal time, which is related to the cosmic time via $\dd t = a \dd\tau$. 

\section{A Minimal Model of Asymmetric Dark Matter}
\label{sec:adm}

Before we discuss the details of the asymmetry generation mechanism, which will follow in Sec.~\ref{sec:asymmetry}, we first introduce the matter content and interactions of the messenger and dark sectors. We consider a minimal asymmetric dark matter model, adopted from Ref.~\cite{Shelton:2010ta}, that is suitable for cogenerating an asymmetry in both the SM and ADM sectors from the same source of $\cal CP$ violation produced during pseudoscalar inflation. 

We introduce a messenger sector that includes two left handed fermions, $L_1, L_2$, which are $\suL$ doublets, and two right handed singlets, $R_1, R_2$, where all of them carry dark lepton global charges, $\dLo,\dLt$, respectively. The dark sector has two Dirac fermions, $X_1,X_2$ that are charged under the dark gauge group $\ud$ and also carry global dark lepton charges $\dLo,\dLt$, respectively. The fermion content is chosen such that it is minimal to cancel all the gauge anomalies of both the SM and dark gauge sector $\ud$ as well as the global Witten anomaly \cite{Witten:1982fp}. The fermionic field content of this minimal ADM model is summarized in Table \ref{tab:adm}, and its Lagrangian is given by 
\be
\mathcal{L}_{\scriptscriptstyle \rm ADM} = 
iL^\dagger_i \bar{\sigma}^\mu D_\mu^L L_i 
+iR^\dagger_i\sigma^\mu D_\mu^R R_i 
+i\bar{X}_i\gamma^\mu D^X_\mu X_i 
- \frac{1}{4} C_{\mu\nu} C^{\mu\nu} 
+\mathcal{L}_{\scriptscriptstyle \rm Yuk}
\com
\ee
where $\bar \sigma^{\mu} = (1, - {\bm \sigma})$,  $\sigma^{\mu} = (1, {\bm \sigma})$, $\{\sigma^i\}$ are Pauli matrices, $\{\gamma^\mu\}$ are Dirac matrices,
\bse
D_\mu^L &=& \partial_\mu + i g_{\rm y} Y_{L_i} {A}_\mu+ig_{\rm w}\frac{\sigma^a}{2}{W}^a_\mu +i\gd C_\mu \com \\
D_\mu^R &=& \partial_\mu + i g_{\rm y} Y_{R_i} {A}_\mu +i\gd C_\mu \com \\
D^X_\mu &=& \partial_\mu + i \gd C_\mu \com
\ese
${A_{\mu}}, {W_\mu}^a, C_{\mu}$ are the $\uy$ hypercharge, $\suL$ weak and $\ud$ dark gauge fields, respectively, and $g_{\rm y}, g_{\rm w}, g_{\D}$ are the corresponding gauge couplings. The Yukawa Lagrangian is given by
\be
\mathcal{L}_{\scriptscriptstyle \rm Yuk} =
 \yLo L_1^\dag H^c R_1 
 + \yLt L_2^\dag H R_2 
+ \yXLo L_1^\dag H X_{1}^\R 
+ \yXLt L_2^\dag H^c X_{2}^\R +  {\rm h.c.} \com
\ee
the Higgs doublet, $H$, and its conjugate 
\be 
H = \left(\begin{array}{c} H^+\\ H^0\end{array}\right) \com
\qquad  
H^c = i\sigma_2 H^\ast=\left(\begin{array}{c} {H^0}^\dag \\ -H^- \end{array}\right) \label{Higgsdoublet} \per
\ee
have hypercharges $Y_H = 1$ and $Y_{H^c} = -1$ , respectively,
\begin{table}[t]
\begin{center}
\renewcommand\arraystretch{1.2} 
\begin{tabular}{|*{9}{c|}}
\hline
\multicolumn{3}{|c|}{Fermions}&\multicolumn{4}{|c|}{Gauge charges}&\multicolumn{2}{|c|}{Global charges} \\\hline
 &field & handedness& $I_3$ & $Y$ &$Q$&$\qd$&~~$\dLo$~~&$\dLt$ \\\hline
\multirow{6}{*}{\rotatebox[origin=c]{90}{Messenger}}
&$L^{u}_1$&left&1/2&1&1&1&1&0 \\\cline{2-9}
&$L^{d}_1$&left&-1/2&1&0&1&1&0 \\\cline{2-9}
&$R_1$&right&0&2&1&1&1&0\\\cline{2-9}
&$L^{u}_2$&left&1/2&-1&0&-1&0&1 \\\cline{2-9}
&$L^{d}_2$&left&-1/2&-1&-1&-1&0&1 \\ \cline{2-9}
&$R_2$&right&0&-2&-1&-1&0&1 \\\hline
\multirow{2}{*}{\rotatebox[origin=c]{90}{DM}} 
&$X_1$&both&0&0&0&1&1&0  \\\cline{2-9}
&$X_2$&both&0&0&0&-1&0&1  \\\hline
\end{tabular}
\end{center}
\caption{Messenger and dark sector fermion content, their chiralities and local and global charges. Weak isospin, $I_3$, hypercharge, $Y$, and electromagnetic charge, $Q$, are the SM electroweak $\suL \times \uy$ charges, whereas $\qd$ is the gauged dark $\ud$ charge. All the fermions have dark lepton-like charges given by $\dLo$ and $\dLt$. We use the convention $Q= I_3 + Y/2$. 
\label{tab:adm}
}
\end{table}

The messenger states
\be
L_i = \left(\begin{array}{c} L_i^u\\ L_i^d\end{array}\right)
\ee
are much more massive than the DM states, $X_i$, hence, they can decay via the following channels:
\be \label{decayLtoX} 
L_1 \longrightarrow X_1^\R + H \com
\qquad
L_2 \longrightarrow X_2^\R + H^c 
\per
\ee
These are the only decays that conserve all the charges in Table \ref{tab:adm}. 
As we will show in Sec.~\ref{sec:asymmetry}, an asymmetry is first generated in the messenger sector, and then gets transferred to the dark sector via the decays given in \eref{decayLtoX}. Once the asymmetry is generated for the right-handed component of $X_i$, the dark gauge interaction $\ud$ equilibrates the left and right handed dark fermions. Since the decay rate
\be\label{GammaLtoX}
\Gamma_{L_i\to X_i^\R + H} \sim \dfrac{1}{8\pi} \yXLi m_\L \com
\ee
is much larger than the Hubble rate $H \sim T^2/M_{\text{pl}}$ for messenger masses of $m_\L \sim 1$ TeV, for instance, we will consider that the asymmetry generated in the messenger sector gets quickly converted into the dark matter states. Hence, we assume that the number densities are related as $n_{X_i}=n_{L_i}$ in what follows.

In addition to the asymmetric component of the ADM, there will be a symmetric part that is thermally produced. In order to annihilate the symmetric component efficiently, we will consider two scenarios, where the $\ud$ dark photons can be massless or massive. We will discuss both cases in detail in Sec.~\ref{sec:annihilation}. We will now present our proposed co-generation mechanism in detail in the following section.

\section{Pseudoscalar Inflation and Asymmetry Generation}
\label{sec:asymmetry}

The asymmetry in both the SM and the messenger sector is generated via the coupling of the hypercharge gauge field to a pseudoscalar inflaton as was studied in Ref.~\cite{Anber:2015yca} for baryogenesis. The Lagrangian that we will consider has the form
\be \label{phi lagrangian}
{\cal L}=\frac{1}{2}(\partial_\mu \Phi)^2-V(\Phi)-\frac{1}{4}{Y}_{\mu\nu}{Y}^{\mu\nu}-\frac{\alpha}{4f}\Phi {Y}_{\mu\nu}\tilde {Y}^{\mu\nu} \com 
\ee
where $\Phi$ is a pseudoscalar inflaton field, $V(\Phi)$ is a flat potential satisfying the slow-roll conditions, ${Y}_{\mu\nu}$ is the hypercharge field strength, $\alpha$ is a dimensionless coupling, and $f$ is the axion constant of dimension mass. 

As the inflaton slow-rolls, it provides a time dependent background to populate the modes of the hypercharge gauge fields due to the dimension-5 coupling given in \eref{phi lagrangian}. Besides, since the inflaton under consideration is a pseudoscalar, it will only lead to over abundance in a given helicity mode of the gauge fields. Hence,  as a result, hypermagnetic fields that are coherent over the horizon scale at the given epoch are produced with net  Chern-Simons density (or magnetic helicity) that breaks ${\cal CP}$ macroscopically \cite{Anber:2006xt} (see also Refs.~\cite{Turner:1987bw,Ratra:1991bn,Dolgov:1993vg,Gasperini:1995dh,Adshead:2016iae}). It has been recently noted in Ref.~\cite{Anber:2015yca} that such hypermagnetic fields can source the baryon asymmetry of the Universe through the chiral anomaly in the Standard Model as all the Sakharov criteria \cite{Sakharov:1967dj} are satisfied in this process (see also 
Refs.~\cite{Joyce:1997uy,Campbell:1992jd,Long:2013tha,Long:2016uez,Giovannini:1997eg,Giovannini:1997gp,Giovannini:1999by,Giovannini:1999wv,Bamba:2006km,Bamba:2007hf,Boyarsky:2011uy,Boyarsky:2012ex,Zadeh:2015oqf,Boyarsky:2015faa,Gorbar:2016qfh,Semikoz:2003qt,Semikoz:2004rr,Semikoz:2005ks,Semikoz:2009ye,Akhmet'ev:2010ba,Dvornikov:2011ey,Semikoz:2012ka,Dvornikov:2012rk,Semikoz:2013xkc,Semikoz:2015wsa,Semikoz:2016lqv,Sabancilar:2013raa,Fujita:2016igl,Kamada:2016eeb,Kamada:2016cnb} 
for the evolution of hypermagnetic fields and their effect on particle asymmetries). It was found that the observed baryon asymmetry can be easily achieved in a generic set of inflaton parameters, i.e., the Hubble rate during inflation $\Hinf$ and the coupling of the inflaton to the hypercharge field, $\alpha$. In this work, we extend this mechanism to include a messenger sector that carries both dark and SM charges so that we can relate the observed dark matter and baryon abundances in the Universe to a single source of ${\cal CP}$ violation generated during inflation.

As the messenger sector fermions carry the SM charges, hence the hypercharge, there will be an accompanying asymmetry in $L_i$ and $R_i$ fermions. Subsequently, the decay of $L_i$ leads to the transfer of this asymmetry into the dark fermions $X_i$. To set our notation, we will briefly go over the chiral anomaly in the SM, review how the helical hypermagnetic fields are generated during inflation, discuss how they evolve in the primordial plasma and finally derive the the kinetic equations governing the evolution of asymmetries in particle species in the SM, messenger and dark sectors. We present our main results at the end of this section.

\subsection{Chiral Anomaly}
\label{sec:chiral anomaly}
Since the SM has chiral fermions, there is a chiral anomaly associated with each species, both gauge and global \cite{'tHooft:1976up}. The gauge anomaly is terminal, but it is cancelled in the SM \cite{'tHooft:1976up}. However, there remains a global anomaly, namely the baryon, $B$, and lepton, $L$, numbers are separately anomalous in the SM so do the additional global dark charges $\dLo,\dLt$ for the messenger sector that we have introduced. To put it in a compact form, each species that are charged under the SM gauge groups exhibit the chiral anomaly given as (see e.g., Ref.~\cite{Anber:2015yca} or appendix of Ref.~\cite{Long:2013tha} for the full set of anomaly equations)
\be \partial_{\mu} j_{\sc f}^{\mu} = \cy^f \frac{\alphay}{16 \pi} Y_{\mu\nu} \tilde Y^{\mu\nu} + \cw^f \frac{\alphaw}{8 \pi} W_{\mu \nu}^a \tilde{W}^{a \, \mu \nu} +\cs^f \frac{\alphas}{8 \pi} G_{\mu \nu}^a \tilde{G}^{a \, \mu \nu} , \label{anomaly eq} 
\ee 
where the coefficients $C_j^f$ are given in Table \ref{tab:coefficients} for all the chiral fermions in the SM and the messenger sector for the corresponding SM gauge groups. $\alpha_j$'s are the fine structure constants of the corresponding SM gauge groups, $\alpha_j=g_j^2/(4\pi)$. 
\begin{table}[h]
\begin{center}
\begin{tabular}{|c|c|c|c|}
\hline
 $f$& $\cy^f$ & $\cw^f$ & $\cs^f$ \\\hline
$\begin{array}{c}q \end{array}$ & $\Nc \Nw \yql^2$ & $\Nc$ & $\Nw$ \\\hline
$\begin{array}{c}\ell \end{array}$ & $\Nw \yll^2$ & $1$ & $0$ \\\hline
$\begin{array}{c}u_\R \end{array}$ & $-\Nc \yur^2$ & $0$ & $-1$ \\\hline 
$\begin{array}{c} d_\R \end{array}$ & $-\Nc \ydr^2$ & $0$ & $-1$ \\\hline
$\begin{array}{c} e_\R \end{array}$ & $-\yer^2$ & $0$ & $0$ \\\hline
$\begin{array}{c} L_i \end{array}$ & $\Nw \yli^2$ & $1$ & $0$ \\\hline
$\begin{array}{c} R_i \end{array}$ & $-\yri^2$ & $0$ & $0$ \\\hline
\end{tabular}
\end{center}
\vspace{-15pt}
\caption{\label{tab:coefficients}
Coefficients $C_j^f$ in \eref{anomaly eq}. The multiplicities $\Nc = 3$ and $\Nw = 2$ take into account the color and weak isospin states of a given family of leptons and quarks, and the SM hypercharges are $\yql = 1/3\com ~\yll = - 1 \com ~\yur = 4/3 \com ~\ydr = - 2/3\com ~\yer = -2$. The charge conjugates $q^c$, $\ell^c$, $u_{\R}^c$, $d_{\R}^c$, $e_{\R}^c$, $L_i^c$ and $R_i^c$ have the same coefficients, $C_j^f$, with all the signs flipped. \label{CoefTable}}
\end{table}

The currents associated with the baryon and lepton numbers in terms of the individual fermionic currents are 
\bse j^\mu_B&=&\frac{1}{3}\sum_{i=1}^3 \l(j^\mu_{q^i}+j^\mu_{u^i_\R}+j^\mu_{d^i_\R}\r) \com
\label{Bcurrent}\\
j^\mu_L&=&\sum_{i=1}^3 \l(j^\mu_{\ell^i}+j^\mu_{e^i_\R} \r) \com 
\ese
and for the both dark lepton like numbers
\be j^\mu_{L_D^i} = j^\mu_{L_i}+j^\mu_{R_i}+j^\mu_{X_i} \per
\ee
We note that all these currents are anomalous, and thus not conserved:
\be 
\frac{1}{N_g}\partial_\mu j^\mu_B =\frac{1}{N_g} \partial_\mu j^\mu_L=\partial_\mu j^\mu_{L_D^i} = \dfrac{\alphaw}{8\pi}W_{\mu\nu}^{a}\tilde{W}^{\mu\nu\,a}-\dfrac{\alphay}{8\pi}Y_{\mu\nu}\tilde{Y}^{\mu\nu},
\label{anomaly}
\ee
where $N_g=3$ is the number of generations in the SM. There is an accidental conserved current, $\partial_\mu j^\mu_{B-L} =0$ in the SM, which has important consequences for the baryon asymmetry of the Universe, namely, baryons can be converted into leptons and vice versa with the selection rule $3 \,\Delta N_L=\Delta N_B$ since any baryon number violation is compensated by a lepton number violation \cite{'tHooft:1976up}. Similarly, $\partial_\mu j^\mu_{L_D^1-L_D^2} =0$ in our setup due to the similarity of the fermionic field content, hence, 
\be\label{messenger baryon ratio}
3 \,\Delta N_{L_D^i}=3 \,\Delta N_L =\Delta N_B.
\ee
Moreover it trivially follows that all the gauge currents \cite{Schwartz:2013pla}
\bse j^\mu_{\rm y} &=& \sum_{\text{particles}} Y_{\psi}\bar{\psi}\gamma^\mu \psi \com
\\\
j^{\mu\,a}_{\rm w} &=& \sum_{\text{left particles}} \bar{\psi}_i \tau^a_{ij}\gamma^\mu \psi_j \com
\\
j^{\mu\,a}_{\rm s} &=& \sum_{\text{quarks}} \bar{\psi}_i \eta^a_{ij}\gamma^\mu \psi_j \com
\\
j^\mu_{ Q_D} &=& \sum_{\text{dark particles}} Q^D_{\psi}\bar{\psi}\gamma^\mu \psi \com
\ese
with $\tau^a_{ij}$, $\eta^a_{ij}$ the generators of $\suL$, $\suc$ respectively, are not anomalous, i.e., $\partial_\mu j^\mu_{\rm y} =\partial_\mu j^{\mu\,a}_{\rm w}=\partial_\mu j^{\mu\,a}_{\rm s}=\partial_\mu j^\mu_{ Q_D}=0$, which ensure that unitarity is not violated. It is well known that the non-Abelian gauge theories have topologically distinct degenerate vacua, transition between which leads to the change of baryon, lepton and dark lepton numbers. However, we stress that $\uy$ sector can also source the chiral anomaly provided that there is a net $Y_{\mu\nu}\tilde{Y}^{\mu\nu}$, e.g., hypermagnetic fields with net helicity. In other words, in the Abelian case, the magnetic helicity is the Abelian Chern-Simon number. We will explain next how such field configurations are produced during inflation.

\subsection{Hypermagnetic Fields from Pseudoscalar Inflation}
\label{sec:field generation}

In this section, we summarize the generation of helical hypermagnetic fields during pseudoscalar inflation to set the notation and to make the paper self contained, see Ref.~\cite{Anber:2006xt} for details. 

The equation of motion for the hypercharge field strength ${Y}_{\mu\nu}$ derived from \eref{phi lagrangian} is 
 \be
 g_{\mu\nu}\nabla^\mu Y^{\nu\rho}=-\frac{\alpha}{f}g_{\mu\nu}(\nabla^\mu \Phi)\tilde{Y}^{\nu\rho} \com
 \ee
where $g_{\mu\nu}$ is the flat FRW metric and $\nabla^\mu$ is the corresponding covariant derivative. 
Using the radiation gauge $A_0=0$, $\bm{\nabla}\cdot \bm{A}=0$, we obtain the equation of motion for the gauge field as 
\begin{eqnarray}\label{main equation of A}
 \l(\pder{^2}{\tau^2}-\nabla^2-\dfrac{\alpha}{f}\pder{\Phi}{\tau}\;\bm{\nabla}\times \r) \bm{A} =0\,,
\end{eqnarray}
where the terms involving ${\bm \nabla} \Phi$ drop out due to the homogeneity of the inflaton field. In this gauge, the hyperelectric and hypermagnetic fields are respectively given by $\bm{E}=-\partial_\tau\bm{A}/a^2$ and $\bm{B}=\bm{\nabla} \times \bm{A}/a^2$, where $a$ is the scale factor in the FRW universe. We promote the vector potential to a quantum operator in the Heisenberg picture
\be 
\bm{\hat{A}}(\tau,\bm{x})=\sum_{\lambda = \pm}\int \dfrac{\dd^3k}{(2\pi)^{3/2}}\l[\bm{\epsilon}_\lambda \hat{a}_\lambda(\bm{k})A_\lambda(\tau,\bm{k})\e{i\bm{k}\cdot \bm{x}}+\bm{\epsilon}^\ast_\lambda \hat{a}_\lambda^\dagger(\bm{k})A_\lambda^\ast(\tau,\bm{k})\e{-i\bm{k}\cdot \bm{x}}\r] ,\label{Aquantized}
\ee
where we used the circular polarization basis $\bm{\epsilon}_\pm $ that obey to the properties $ \bm{k}\cdot \bm{\epsilon}_\pm=0$, and $\bm{k}\times \bm{\epsilon}_\pm=\mp i|\bm{k}|\bm{\epsilon}_\pm$, such that $|\bm{\epsilon}_\pm|^2=1$.
Defining a parameter
\be 
\xi=\frac{\alpha\dot{\phi}_0}{2f\Hinf} \com
\ee
and using the fact that $a(t)=\e{\Hinf t}$ during inflation we obtain
\be \label{potinflation}
\pder{^2A_\pm}{\tau^2} +k\l(k\mp \frac{2\xi}{\tau}\r)A_\pm=0 \per 
\ee
We distinguish three cases that lead to solutions with different asymptotic behaviors. At early times, when $|k\tau| \gg |2\xi|$, the solution is a vacuum mode, hence a free wave $A_\pm=\e{-ik_0\tau}/\sqrt{2}$. When $|k\tau| \sim |2\xi|$ the field develops an instability. Depending on the sign of $\xi$, either $A_+$ or $A_-$ modes will be amplified ($\xi \gtrless 0 \Leftrightarrow A_\pm$ amplified). In the limit $|k\tau| \ll |2\xi|$ the solution for the growing mode is given by \cite{Anber:2006xt} 
\begin{eqnarray}\label{produced field}
A_{\pm}\cong \frac{1}{\sqrt{2k}}\left(\frac{k}{2\xi~ a(\tau) \Hinf}\right)^{1/4}\e{\pi \xi -2\sqrt{2\xi k/[a(\tau)\Hinf] }} \per
\label{latetime}
\end{eqnarray}
whereas the other mode is exponentially suppressed. Note that due to the $\e{\pi \xi}$ factor, the gauge potential grows tremendously for moderate values of $\xi\geqslant1$.

\subsection{Evolution of Hypermagnetic Fields in Plasma}
\label{sec:field evolution}
We assume instant reheating so that immediately after inflation the Universe becomes filled with a plasma of relativistic particles. Therefore, the evolution of hypermagnetic and hyperelectric fields are governed by the relevant magnetohydrodynamics (MHD) equations \cite{Giovannini:1997eg}
\bse
\pder{\bm{B}}{t}&=&-\bm{\nabla}\times\bm{E}\label{MHD1} \com
\\ 
\pder{\bm{E}}{t}+\bm{J}&=&\bm{\nabla}\times\bm{B}\label{MHD2} \com
\\ 
\bm{\nabla}\cdot \bm{B}&=&0 \com
\\ 
\bm{\nabla}\cdot \bm{E}&=&\rho \com
\\
\bm{\nabla}\cdot \bm{J}&=&0 \com
\\ 
\bm{J}&=&\sigma(\bm{E}+\bm{v}\times\bm{B}) \com
\label{MHD6}
\ese
 where ${\bm v}$ is the plasma fluid velocity and  $\sigma \simeq100 T$ is the hypercharge conductivity \cite{Arnold:2000dr}. 
 Assuming a neutral plasma, $\rho =0$, with sufficiently slowly varying hyperelectric field such that $\partial_t \bm{E}=0$, and combining equations (\ref{MHD1}), (\ref{MHD2}) and (\ref{MHD6}), we obtain the evolution equations for the hypermagnetic fields
 \bse 
 \pder{\bm{B}}{t}&=&\bm{\nabla}\times(\bm{v}\times\bm{B})+\dfrac{1}{\sigma}\nabla^2\bm{B} \label{MHDcomb1}
 \com \\
 \bm{E}&=& \dfrac{1}{\sigma} (\bm{\nabla}\times\bm{B})-\bm{v}\times\bm{B}\label{MHDcomb2} 
 \per
 \ese
 The former equation states that the time evolution of the hypermagnetic field depends on an advection term and a dissipation term when the hyperconductivity is finite. The Reynolds number $\mathcal{R}$ is defined as the ratio of these two terms in the Fourier space 
\be 
\mathcal{R} = \dfrac{v \sigma}{k_p} \com
\ee 
where $k_p$ is the last mode that exits the horizon after inflation
\be 
k_p\simeq \dfrac{\Hinf}{\xi}\dfrac{T}{T_{\rm{rh}}} \per
\ee
For $\mathcal{R}<1$ the hypermagnetic field will quickly dissipate as the dissipation term dominates whereas for $\mathcal{R}>1$ a turbulent flow will be generated, and hence, the magnetic field will be sustained. 
Assuming instant reheating, the reheating temperature can be estimated as
\be 
T_{\rm{rh}} \simeq \frac{1}{4}\sqrt{\Mpl \Hinf} \com
\ee
and thus,
\be 
\mathcal{R} = 25 v \xi \sqrt{\dfrac{\Mpl}{\hinf}}, 
\ee
which is much bigger than unity for velocities \cite{Anber:2015yca}
\be 
v > \dfrac{10^{-5}}{\xi} \sqrt{\dfrac{\Hinf}{10^{14}\text{ GeV}}} \per
\ee
We therefore consider the plasma as turbulent in what follows. In the next section, we show that this condition leads to conservation of helicity of the maximally helical hypermagnetic fields generated during inflation.

\subsection{Hypermagnetic Helicity}
\label{sec:helicity}
Hypermagnetic helicity feeds into the chiral anomaly and eventually sources the baryon and dark matter asymmetries. We can deduce the useful relation from \eref{MHDcomb2} 
\be 
\edotb = \frac{1}{\sigma} {\bm B} \cdot {\bm \nabla} \times {\bm B} \label{BE} \com
\ee
where we used ${\bm B}\cdot {\bm v} \times {\bm B} =0$.

The magnetic helicity is defined as
\be 
\mathcal{H}=\int \dd^3x ~\bm{A}\cdot \bm{B} \per
\ee
Using the MDH equations and relation (\ref{BE}), we obtain the rate of change of the spatially averaged helicity density as 
\be \label{hdotB} 
\pder{h}{t}=-\lim_{V\to \infty}\dfrac{2}{\sigma V}\int_V  \dd^3x~ \bm{B}\cdot \bm{\nabla}\times\bm{B} \per
\ee
Using \eref{Aquantized}, we obtain the spatially averaged quantity of interest
\begin{eqnarray}
\label{B curl B}
\langle {\bm B} \cdot {\bm \nabla} \times {\bm B} \rangle_{\rm inf}=\frac{1}{a^5} \int \frac{\dd^3k|\boldsymbol k|^3}{(2\pi)^3} \left(|A_+|^2-|A_-|^2\right)\,,
\end{eqnarray}
where the integral is over the comoving momenta ${\bm k}$. Note that only one of the modes $A_\pm$ is amplified as shown in the previous section. Thus, the produced field has maximal helicity. After setting one of the modes to zero and using \eref{latetime}, we obtain\footnote{We note that the the produced hypermagnetic fields are maximally helical saturating the realizability condition $h_{\rm M}(k) \leqslant 2 e_{\rm M}(k)/k$ (see e.g., Ref.~\cite{Field:1998hi}). Here, for instance for a given helicity mode, say $A_{+}$, the following relations are always satisfied: $\int {\rm d} k ~h_{\rm M}(k) \equiv {\frac{1}{V}}\int_{V} {\rm d}^3 x~ \moy{\bm{A}\cdot \bm{B}} = \int {\rm d} k~ k^3 ~|A_+|^2$ whereas $\int {\rm d} k~ e_{\rm M}(k) \equiv \frac{1}{V} \int_{V} {\rm d}^3 x~\frac{1}{2}~ \moy{\bm{B}^2} =\frac{1}{2} \int {\rm d} k~ k^4 ~|A_+|^2$.} \cite{Anber:2015yca}
\be  
\moy{\bm{B}\cdot\bm{\nabla}\times\bm{B}}=\; \pm I \dfrac{H^5\e{2\pi\xi}}{\xi^6} \com
\ee
where $I=6.848\cdot 10^{-4}$ and the overall sign depends on the choice of the mode $A_{\pm}$, respectively. Here, we cut off the integral at $k_c\simeq 2\xi H a(\tau)$ in order to be in the range of validity of the expression for $A_\pm$. When performing the integral, we ignored residual terms that are proportional to $\xi^{-1}\e{-8\xi}$ as they are exponentially suppressed since $\xi \gg |k\tau|$.
Therefore, at the end of inflation, the change of helicity finally reads \cite{Anber:2015yca}
\be 
\pder{h}{t}=\mp2I\dfrac{\e{2\pi\xi}}{\sigma \xi^6}\l(\dfrac{\Hinf}{a}
 \r)^5, \label{helicity_final}
 \ee
Here, cosmological redshift has been taking into account and the scale factor has been normalized such that it is one at the end of inflation. We note that to generate baryons rather than antibaryons, the negative sign has to be chosen, corresponding to the mode $A_+$ as we will see in the next section.

\subsection{Kinetic Equations}
\label{sec:boltzmann}

The set of kinetic equations is found by integrating the anomaly equations (\ref{anomaly eq}) over spacetime.
The number density of a given particle species in terms of a current reads
\be \label{Ndensity}
n_i=\lim_{V\to \infty}\dfrac{1}{V}\int_V\dd^3x\; j^0_i \com
\ee
and defining the asymmetry parameter of a given species as
\be
\eta_f = \frac{n_f-n_{\bar{f}}}{s} \com
\ee
where, $s$ is the entropy density. The relevant asymmetry parameters for the SM and messenger sector fermions are
\bse 
\etaql &=& \frac{1}{6s}  \NgSM \Nw \Nc \muql T^2 \label{eta1}
\com \\
\etall &=& \frac{1}{6s}  \NgSM \Nw \mull T^2 
\com \\
\etaur &=& \frac{1}{6s}  \NgSM \Nc \muur T^2 
\com \\
\etadr &=& \frac{1}{6s}  \NgSM \Nc \mudr T^2 
\com \\
\etaer &=& \frac{1}{6s}  \NgSM \muer T^2 
\com\\
\etaL &=& \frac{1}{6s}  \NgDM \Nw \muL T^2 \com
\\
\etaR &=& \frac{1}{6s}  \NgDM \muR T^2 \com \label{eta7}
\ese
where $\NgSM =3$ and $\NgDM =2$ are the multiplicity factors for the SM and messenger sector families, respectively.

Upon integrating and thermally averaging the right hand side of the anomaly equations (\ref{anomaly eq}), we obtain three contributions. The first contribution comes from the hypercharge sector through the term $Y_{\mu\nu}\tilde{Y}^{\mu\nu}= -4\,\edotb$ which brings the rate of change of helicity density as we derived in Eqs.~(\ref{BE}), (\ref{hdotB}) and (\ref{helicity_final}). The other two contributions come from the $\suL$ and $\suc$ sphalerons, i.e., the weak and the strong sphalerons, respectively. The weak sphalerons relax the baryon+lepton number charge of the fermions charged under $\suL$ whereas the strong sphalerons relax the chiral charge of the quarks charged under $\suc$ \cite{Kuzmin:1985mm,Khlebnikov:1988sr,Rubakov:1996vz}. We note that since sphalerons act on a global level, relaxing global charges, we defined the asymmetry parameters as a sum over all internal degrees of freedom (spin, color, isopsin and family). Finally, the kinetic equation\footnote{Here we neglect both the Yukawa terms and the chiral magnetic effect, which can change the final values of the asymmetry parameters slightly. See, e.g., Ref.~\cite{Fujita:2016igl} and Ref.~\cite{Kamada:2016eeb}, where the Yukawa terms and the chiral magnetic effect are taken into account for baryogenesis, respectively.} corresponding to the seven asymmetry parameters given in Eqs.~(\ref{eta1})-(\ref{eta7}) is (see Ref.~\cite{Anber:2015yca})
\be \label{Boltzmannset}
\frac{\partial \eta_f}{\partial t} = \cy^f \dfrac{\alphay}{ 4\pi s} \pder{h}{t} - \cw^f \gammaw (\etaql+\etall+\etaL) - \cs^f \gammas (\etaql - \etaur - \etadr) \per
\ee

The coefficients $C_j^f$ are given by the Table \ref{CoefTable}. In \eref{Boltzmannset},  $\gammaw = 25 \alphaw^5T$ \cite{Moore:1997sn} and $\gammas = 100 \alphas^5T$ \cite{Moore:1997im} are the weak and strong sphaleron rates per unit time, respectively.
Notice that this set of equations respect the Sakharov conditions \cite{Sakharov:1967dj} since: 1) the anomalous $B/L/L_D^i$ currents provide a $B/L/L_D^i$ number violation; 2) the term containing $\dot{h}$ has different sign for different chiralities hence breaks $\cal C$/$\cal CP$; 3) the $\dot{h}$ term is a source term (external field produced during inflation) and hence describes an out of equilibrium process.

Since we add new species to the Standard Model, the number of relativistic degrees of freedom increases:
\be \label{gchange} 
g_\ast=g_\ast^{\text{SM}}+g_\ast^{\text{M}}+g_\ast^{\text{DM}}=106.75+\frac{7}{8}\cdot 12+\frac{7}{8}\cdot 8 + 2 =126.25 \com
\ee
where we also considered a massless dark photon corresponding to the dark $\ud$ gauge field.

It is more convenient to express \eref{Boltzmannset} in terms of a dimensionless variable $x$ defined as
\bse 
x&=&D\dfrac{M_{\text{pl}}}{T} \com
\\ D &=& \sqrt{\dfrac{45}{4\pi^3g_\ast}} \per
\ese
As soon as inflation ends, the radiation dominated era begins, and thus, we have the relation $H=1/(2t)=T^2/(M_{pl} D)$. Performing this change of variable in \eref{Boltzmannset} yields our final master equation
\be \frac{\partial \eta_f}{\partial x} = -\cy^f \gammay - \cw^f \gammaws (\etaql+\etall+\etaL)- \cs^f \gammass (\etaql - \etaur - \etadr) \com
\ee
with
$\gammay = \frac{{\cal I} e^{2\pi \xi} \alphay }{\xi^6 \sqrt{D}}\frac{T}{\sigma} \left(\frac{\hinf}{\Mpl}\right)^{5/2}$ and $\gamma_{\rm w/s} =\frac{\Gamma_{\rm w/s}}{T}$. This equation is much more convenient to solve since the system of equations becomes just a set of first order differential equations with constant coefficients.

This set of kinetic equations is valid from the end of inflation until the weak sphalerons shut off, at temperature $T = \frac{8\pi v}{\sqrt{4\pi \alphaw}} \simeq 10$ TeV, where $v=246$ GeV is the Higgs vacuum expectation value. For simplicity we assumed that the $\gammay$ source shuts off at the same temperature even if it contributes to the evolution of $\eta_f$ until electroweak phase transition, $T_\text{EW} \simeq 160$ GeV.

\subsection{Results}
\label{sec:results}

\begin{figure}
\begin{center}
\includegraphics[width = 10cm]{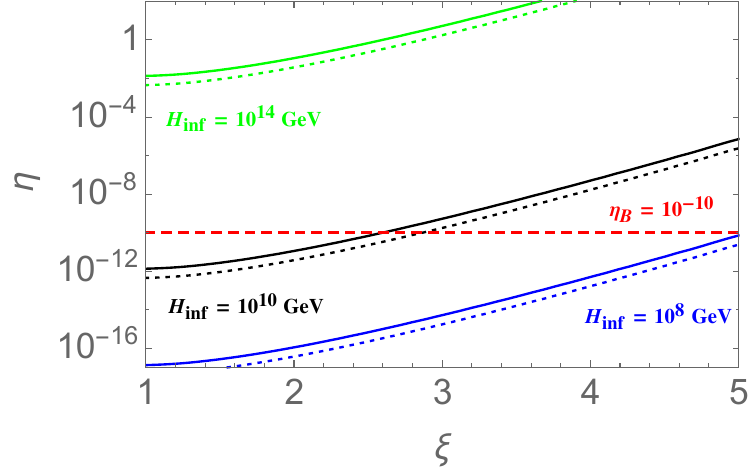}
 \caption{Numerical solutions of the kinetic equations given by \eref{Boltzmannset}. The lines show the result for the baryon asymmetry and the dotted one corresponds to the asymmetry in one family of DM. The red dashed line shows the observed value of the baryon asymmetry $\eta_B\simeq10^{-10}$ which can be achieved for instance with $\xi =1$ and $H_{\text{inf}}=5.6\cdot 10^{10}$ GeV. The messenger sector asymmetry is $\eta_L=\eta_B/3$ for all values of $H_{\text{inf}}$ and $\xi$ as also can be obtained from the relation \eref{messenger baryon ratio}.}
 \label{VM_Boltz_results}
\end{center}
\end{figure}

\begin{figure}
\begin{center}
\includegraphics[width = 8cm]{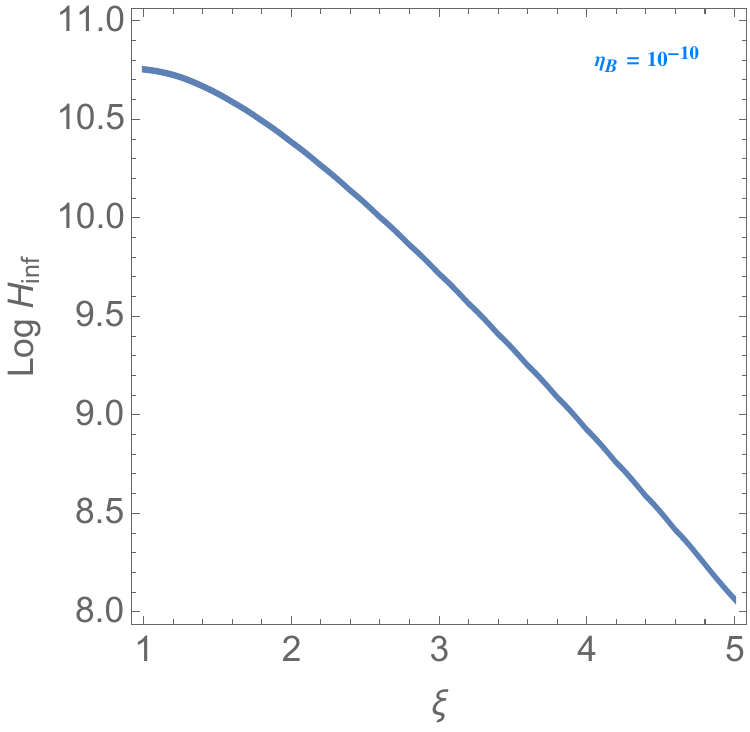}
 \caption{Numerical solutions of the kinetic equations given by \eref{Boltzmannset}. This plot shows the values of the parameters $H_{\text{inf}}$ and $\xi$ in order to have the observed baryon asymmetry  $\eta_B\simeq10^{-10}$ with $\eta_L = \eta_B/3$.}
\label{VM_Boltz_results2}
\end{center}
\end{figure}

The solution of the system of equations (\ref{Boltzmannset}) allows us to obtain the parameter space for the inflation parameters, namely, $\Hinf$ and $\xi$, in order to produce the observed value of the baryon asymmetry $\eta_B=1\times 10^{-10}$ \cite{Ade:2015xua}. We assume that initially all the asymmetry parameters given by Eqs.~(\ref{eta1})-(\ref{eta7}) are zero. We show the parameter space in Figures \ref{VM_Boltz_results} and \ref{VM_Boltz_results2} for $1\leqslant \xi \leqslant 5$ and $10^8 ~{\rm GeV}\leqslant\hinf\leqslant 10^{14}~ {\rm GeV}$.
Recall that only for $\xi>1$ the analytic solution of the mode is given by \eref{latetime} that we have used in our calculation\footnote{Note that there is a constraint on the parameter $\xi \lesssim 4$  from non-Gaussianities caused by the hypermagnetic fields, see, e.g., Refs.\cite{Ferreira:2014zia,Ferreira:2015omg}.}. 

The solution also provides a relation for any inflation parameters between the asymmetry parameters of the SM and messenger sectors: $\eta_B=3\eta_{L}$ [see \eref{messenger baryon ratio}]. Not surprisingly, because the equations for dark lepton number 1 and 2 are identical, their asymmetry parameters are equal: $\eta_{L_1}=\eta_{L_2}$. Since we take $n_L=n_X$ it is possible to compute the typical mass of the DM candidates as follows.
Today neither the dark matter nor the baryonic matter is relativistic: $ \rho_i=m_in_i=m_i\eta_is_0 = \Omega_i \rho_c$, where $s_0$ is the actual entropy density. This yields the relation 
\be 
\dfrac{\Omega_{X_i}}{\Omega_b}=\dfrac{m_{X_i}}{m_p}\dfrac{\eta_{X_i}}{\eta_B} \com
\ee
for each dark matter particle $X_i$, where $m_p=938.73$ MeV is the proton mass. Since observation cannot distinguish $\Omega_{X_1}$ and $\Omega_{X_2}$ which sum up to $\Omega_{DM}$, we compute the equivalent DM mass by performing a sum on both dark species contribution $m_{DM}=\sum m_{X_i}$ and we get:
\be m_{DM} = m_p \dfrac{\eta_B}{\Omega_b} \sum_i \dfrac{\Omega_{X_i}}{\eta_{L_i}}=m_p\dfrac{\Omega_{DM}}{\Omega_b}\dfrac{\eta_B}{\eta_{L_{1,2}}} \simeq 15 \text{ GeV,}\ee which is in the range of the allowed values \cite{Petraki:2013wwa}. This is the total mass of the DM particles and depending on whether one of them is lighter than the other leads to the dominant component mass in the range of $7-15$ GeV. In other words, if $m_{X_1}\sim m_{X_2} $, we predict the DM mass to be around $7$ GeV whereas for $m_{X_1} \gg m_{X_2}$ (or the other way around), the DM mass is predicted to be around $15$ GeV.

Now that we have successfully generated an asymmetry in the both the SM and the dark matter sectors, we will turn our attention into getting rid of the possible thermal symmetric component of dark matter. 

\section{Annihilating the Symmetric Component of Dark Matter}
\label{sec:annihilation}

There are three possibilities in order to annihilate the symmetric part of the DM: 1) the DM annihilates into SM states, 2) it annihilates into messenger states that eventually decay to SM states, 3) it annihilates into dark radiation.
The two first possibilities are forbidden by our model since the DM candidates cannot annihilate into the SM because they do not carry any of the SM charges. Besides their annihilation into the other dark sector species (the messengers of our model) is not efficient enough since they are the lightest. Therefore the only remaining possibility is direct annihilation into dark radiation, that is why we added a gauge interaction between DM states in the first place when we introduced our model in Section \ref{sec:adm}. 

We consider a $\ud$ gauge group whose mediator is the dark photon $\gamma_D$. There are now two cases: either $\gamma_D$ is massless, and we have an unbroken $\ud$ or $\ud$ is broken, and thus, $\gamma_D$ is massive. In the first case the $\gamma_D$ production increases the radiation component in the primordial plasma which affects the big bang nucleosynthesis (BBN) and the cosmic microwave background (CMB). Alternatively, we can take a massive $\ud$, in which case the $\ud$ group is spontaneously broken and $\gamma_D$ can mix kinematically with $\uy$ photon and decay into SM states. This case is safer from cosmological constraints as we will see next. We shall explore these two cases in detail in the next subsections.

As we argued briefly, aside from the asymmetry generation that leads to the observed DM abundance, we have a thermal production of messenger particles with
\be 
n_L= \dfrac{3\zeta(3)}{4\pi^2}g_LT^3 \per
\ee
The symmetric part of the messengers then decay very efficiently to $X_i$ according to \eref{decayLtoX}. DM annihilates into dark photons with cross-section (see e.g., section 4.2 of \cite{Zurek:2013wia})
\be 
\moy{\sigma v} \simeq \frac{\pi \alpha_D^2}{m_{X_i}^2} \simeq2\cdot 10^{-8} \l(\dfrac{g_D}{0.1}\r)^4\l(\dfrac{10 \text{ GeV}}{m_{X_i}}\r)^2 \text{ GeV}^{-2} \per
\ee
The annihilation rate is then simply $\Gamma_{X\bar{X}\to\gamma_D\gamma_D} = n \moy{\sigma v} \gg H$. Hence the symmetric part of the messenger and dark sectors annihilate quickly into dark photons within a Hubble time. Next, we will discuss the massive and massless $\gamma_D$ cases separately. 

\subsection{Massless $\gamma_D$}

It is possible to find out the actual temperature and density of the relic $\gamma_D$ as it is done usually for neutrinos. Using entropy conservation we obtain a relation between the visible and dark sectors after their decoupling that occur at $T \sim m_{X} \sim 10$ GeV 
\be 
\frac{g_VT_V^3}{g_DT_D^3}= \frac{g_V^{dec}}{g_D^{dec}} \com
\ee
where $V$ (respectively $D$) denote the visible (dark) photon. Referring to Table \ref{tab:adm}, we find that
\bse 
g_D^{dec} &=& 2 \cdot 2 \cdot 2 \cdot \frac{7}{8}+ 2= 9  \com
\\g_D &=& 0 \cdot \frac{7}{8}+ 2 = 2 \com
\ese
since at V-D decoupling there are both $X_1$, $X_2$ and their antiparticles and one massless $\gamma_D$. $X_1$ and $X_2$ are not relativistic in the current epoch so we do not count them here, and there is still one massless $\gamma_D$. At $T \sim 10$ GeV, the SM plasma contains every particle except the top quark, the three $W$ bosons, all the Higgs and all the messenger particles. Thus, the corresponding effective number of relativistic degrees of freedom is $g_V^{dec}=86.25$.
We find the temperature of dark photons in the current epoch as
 \be 
 T_{\gamma_D}=\l(\frac{g_D^{dec}}{g_V^{dec}}\frac{g_V}{g_D}\r)^{1/3}T_\gamma 
 = \l( \frac{9}{86.25}\frac{2}{2} \r)^{1/3}T_\gamma  \label{photonsTemp} \per
 \ee
A good measure of extra relativistic degrees of freedom is the effective number of relativistic neutrino species defined as the ratio of energy density of one neutrino species (1 left handed neutrino + 1 right handed antineutrino so there is a factor 2)
\be N_{\text{eff}}^\rho= \frac{\rho}{\rho_\nu} \com
\ee
where we have 
\bse 
\rho_\nu &=& 2\cdot \frac{7\pi^2}{240}T_\nu^4 
\com\\ 
\rho_{\gamma_D} &=& 2\cdot\frac{\pi^2}{30} T_{\gamma_D}^4 \com
\ese
and 
\be T_\nu = \l(\frac{4}{11} \r)^{1/3} T_\gamma \per 
\ee
Therefore, 
\be 
\Delta N_{\text{eff}}^{\gamma_D}=\frac{\rho_{\gamma_D}}{\rho_\nu}=\frac{8}{7}\l(\frac{11}{4} \r)^{4/3}\l(\frac{g_D^{dec}}{g_V^{dec}}\frac{g_V}{g_D}\r)^{4/3}=0.22  \per
\ee
This is smaller than the maximum allowed value $\Delta N_{\text{eff}}=0.334$ by the Planck collaboration \cite{Ade:2015xua} so the massless case is marginally allowed by the cosmological observations. The value can be lowered by two mechanisms: increasing the degrees of freedom in the dark sector or increase the V-D decoupling temperature, which will increase $g_V^{dec}$.

\subsection{Massive $\gamma_D$}

In order to give $\gamma_D$ a mass we must add a dark Higgs field in the model to break $\ud$, that is a complex scalar field $\varphi_D$ with a dark charge $q_D$. The Lagrangian is \cite{Petraki:2014uza}
\be 
\mathcal{L}_{\varphi_D}= D_\mu \varphi_D^* D^\mu \varphi_D - \lambda \l(|\varphi_D|^2 - v_D^2 \r)^2 \com
\ee
with $D_\mu= \partial_\mu + i \gd C_\mu$ and where $v_D$ is the vacuum expectation value of $\varphi_D$. The physical mass of a particle is roughly given by the product of the Higgs VEV and the coupling constant. Since the dark photon mass is 
\be 
M_{\gamma_D}=\sqrt{2}q_Dgv_D \com
\ee 
which we take to be around $100 \text{ MeV}$ as an example, it implies a smaller Higgs VEV and a smaller dark Higgs mass, $m_{\varphi_D}=2\sqrt{\lambda} v_D$. 
The dark Higgs mass is much smaller than the visible Higgs mass, allowing the latter to decay in the former via the term
\be 
\mathcal{L}_{\varphi_DH}= \lambda_{\varphi_DH}|\varphi_D|^2|H|^2 \com
\ee
which turns out to yield a negligibly small contribution.

The kinetic mixing is given by the effective Lagrangian
\be \mathcal{L}_{\text{mix}}= \dfrac{\varepsilon}{2}Y_{\mu \nu}C^{\mu \nu}, \ee
where $C^{\mu \nu}$ is the field strength associated to $\ud$. The kinetic mixing of $\uy$ and $\ud$ is given by the parameter $\epsilon$ and reads \cite{Holdom:1985ag}
\be 
\epsilon \sim \frac{g_Yg_D}{16 \pi^2}\ln{\frac{M_L^+}{M_L^-}} \com
\ee
where $M_L^+$ and $M_L^-$ are respectively the higher and lower mass of the different $L$ messenger states. This simply comes from a 1-loop diagram with messenger fermions in the loop. The logarithmic factor is typically of the order one, hence
\be 
\epsilon \sim 6 \cdot 10^{-3} \l(\frac{g_Y}{0.1}\r)\l(\frac{g_D}{0.1}\r) \per
\ee
Then, the $\gamma_D$ decay rate is \cite{Batell:2009yf,Petraki:2014uza}
\be 
\Gamma_{\gamma_D \to l \bar{l}} = \frac{\epsilon^2 \alpha_Y}{3}M_{\gamma_D} \sqrt{1-\frac{4m_l^2}{M_{\gamma_D}^2}}\l(1+\frac{2m_l^2}{M_{\gamma_D}^2} \r) \com
\ee
for the SM leptons and 
\be  
\Gamma_{\gamma_D \to q \bar{q}}=\l. \frac{\sigma_{e^+e^-\to q \bar{q}}}{\sigma_{e^+e^-\to \mu^+ \mu^-}}\r|_{s=M_{\gamma_D}^2} ~ \Gamma_{\gamma_D \to l \bar{l}}  \com
\ee
for hadrons. Of course, these decay channels are allowed only if $M_{\gamma_D}>2m_f$. The decay rate for an electron-positron channel is
\be \Gamma_{\gamma_D \to e^+ e^-} \simeq 1.2 \cdot 10^{-7} \l(\frac{\epsilon}{6 \cdot10^{-3}}\r)^2\l(\frac{M_{\gamma_D}}{100 \text{ MeV}}\r) \text{ GeV}. \ee
or $\Gamma = 1.82 \cdot 10^{17}$ s$^{-1}$, exceeding the Hubble rate by several orders of magnitude. Thus, the massive dark photon can efficiently be converted into the SM species, and hence, the symmetric component of the DM is removed successfully. For $M_{\gamma_D}<210$ MeV,  this is the only allowed channel. When $M_{\gamma_D}$ increases and it allows more channels (say $n$), hence quicker decay, the total decay rate can be parametrized as 
\be 
\Gamma_d = n\cdot \Gamma_{\gamma_D \to f \bar{f}} \com
\ee
unless we are at a threshold of pair production.

\section{Summary and Discussion}
\label{sec:conclusion}

We proposed a new mechanism to generate asymmetric dark matter and the baryon asymmetry of the Universe via the same source of $\cal CP$ violation that occurs during inflation. The coupling of the inflaton to the SM hypercharge gauge fields via the dimension five operator $(\alpha/f)\Phi F_{\mu\nu} {\tilde F}^{\mu\nu}$ leads to generation of coherent hypermagnetic fields with maximal helicity, which in turn source the chiral anomaly in the SM and yield the desired asymmetries in both the SM and DM sectors. We showed that for a wide range of inflationary parameters, $\Hinf$ and $\xi$, the observed BAU and DM abundances can be achieved. In the minimal ADM model we considered, we found that the DM mass is in the range of $m_{X}\sim 7-15$ GeV, depending on whether they have comparable masses or one of the two DM species is relatively lighter, respectively. The DM mass can take a different range of values if the minimal ADM field content is extended as this will affect the ratio between the BAU and the DM asymmetry parameters. 

We also gave two scenarios for annihilating the symmetric part of the ADM. By coupling the DM fermions to a $\ud$ gauge field, the symmetric part can be efficiently annihilated into the dark photons, $\gamma_D$. In the first scenario we considered, $\ud$ is unbroken, and hence $\gamma_D$ is massless and contributes to the relativistic degree of freedom today. We found that the contribution of $\gamma_D$, $\Delta N_{\rm eff} =0.22 $, is within the allowed range of $N_{\rm eff} = 3.15\pm 0.23$ provided by the Planck collaboration \cite{Ade:2015xua}. However, we note that as the constraint on $\Delta N_{\rm eff}$ continues to improve, this scenario might become problematic. One way out is to increase the field content in the ADM sector to dilute the dark radiation component. In the second scenario that we have considered for annihilating the symmetric part of the ADM, the $\ud$ is broken, hence the dark photons are massive. These photons decay into the SM species via a gauge kinetic mixing efficiently. Hence, this is a safer route to annihilate the symmetric component of ADM and is free from the constraints.  

Since in the minimal ADM model we considered there are two DM candidates carrying two different quantum numbers, there is a possibility for them to combine and form dark Hydrogen-like atoms.  For the massless $\gamma_D$ case, dark atoms can form provided that \cite{CyrRacine:2012fz}
\be 
\dfrac{\alpha_D^6}{\cal R}\l(\dfrac{\Omega_{DM}h^2}{0.11}\r)\l(\dfrac{1\text{ GeV}}{m_{DM}-B_D}\r)\l(\dfrac{1\text{ keV}}{B_D}\r)\geqslant1.5\cdot10^{-16} \com
\ee
with
\be 
{\cal R} = \l.\frac{T_{\gamma_D}}{T_\gamma}\r|_{z=0} \com
\ee
and $B_D$ the binding energy of the dark Hydrogen-like atom. Taking $\Omega_{DM}h^2=0.11$, $B_D=1$ keV, $\gd=0.1$ and our value of ${\cal R} = 0.47$ [see \eref{photonsTemp}], we conclude that this condition is not reached for our value of $m_{DM} =7-15$ GeV. For the massive $\gamma_D$ case, dark atoms can form provided that \cite{Petraki:2014uza}
\be 
m_{\gamma_D}< \alpha_D \dfrac{m_{X_1}m_{X_2}}{m_{DM}} \com
\ee
For $m_{\gamma_D} = 100$ MeV and $g_D=0.1$ this condition cannot be respected either for the range of $m_{DM}$ that we have.
Therefore, in our setup the dark atoms cannot form.

We would also like to point out that the model of ADM that we considered with the dark $\ud$ interaction is compatible with the observation of haloes and subhaloes in the galaxies, namely, addressing the the so-called ``missing-satellite problem" as was discussed extensively in Ref.~\cite{CyrRacine:2012fz}. In a nutshell, this can be understood as follows. The interaction between the DM mediated by a massless or light force carrier reproduces the large-scale structure of the universe while suppressing the formation of structure at smaller scales \cite{Petraki:2014uza}.

Finally, we note that in the simplest version of the natural inflation model \cite{Freese:1990rb} that we considered as an example to study the asymmetry generation in ADM and baryons, either a curvaton field \cite{Lyth:2001nq} is needed to explain the observed density perturbations or some other dynamical mechanism is needed to be implemented (see e.g., string theory inspired models \cite{Enqvist:2001zp,Dimopoulos:2005ac}). To achieve the observed BAU and hence the DM abundance, we require $\Hinf$ to be not so large as can be read off from our Figures \ref{VM_Boltz_results} and \ref{VM_Boltz_results2}. A model of inflation that has $\Hinf \leqslant 5.6 \cdot 10^{10}$ GeV, $\xi \gtrsim 1$ and that also explains observed amplitude of the density perturbations leads to a complete picture of the early Universe, namely, solving all the initial conditions problems including the BAU and DM abundances in a single framework. The interrelations between our proposed mechanism and various models of inflation that satisfy these criteria remain to be studied.

\acknowledgments

We would like to thank Mohamed Anber, Kohei Kamada and Mikhail Shaposhnikov for discussions.
ES is supported by the Swiss National Science Foundation and Alexander von Humboldt Foundation. YC is supported by the Institute of Physics at Ecole Polytechnique F\'ed\'erale de Lausanne. 

\bibliographystyle{JHEP}
\bibliography{adm_refs}

\end{document}